\documentclass{revtex4}

\usepackage{amsmath}    
\usepackage{graphicx}   
\usepackage[usenames,dvipsnames]{xcolor}

\begin{document}
 
\newcommand{\ket}[1]{|#1\rangle}
\newcommand{\bra}[1]{\langle#1|}
\newcommand{\overl}[2]{\langle#1|#2\rangle}
\newcommand{\ph}{|\phi\rangle}
\newcommand{\ps}{|\psi\rangle}

\newcommand{\0}{|0\rangle}
\newcommand{\1}{|1\rangle}

\title{A SWAP gate for qudits}

\author{Juan Carlos Garcia-Escartin}

\affiliation{Universidad de Valladolid, Dpto. Teor\'ia de la Se\~{n}al e Ing. Telem\'atica, Paseo Bel\'en n$^o$ 15, 47011 Valladolid, Spain}
\email{juagar@tel.uva.es}   
\author{Pedro Chamorro-Posada}
\affiliation{Universidad de Valladolid, Dpto. Teor\'ia de la Se\~{n}al e Ing. Telem\'atica, Paseo Bel\'en n$^o$ 15, 47011 Valladolid, Spain}
\date{\today}

\begin{abstract}
We present a quantum SWAP gate valid for quantum systems of an arbitrary dimension. The gate generalizes the CNOT implementation of the SWAP gate for qubits and keeps its most important properties, like symmetry and simplicity. We only use three copies of the same controlled qudit gate. This gate can be built with two standard higher-dimensional operations, the Quantum Fourier Transform and the $d$-dimensional version of the $C\!Z$ gate. 
\end{abstract}
\maketitle

\section{Introduction}
\label{intro}
Quantum information protocols and quantum algorithms are usually designed for a group of two level quantum systems (qubits). With $n$ qubits, we can access Hilbert spaces of dimension $d=2^n$. However, in some occasions it is a good idea to look into systems with an arbitrary dimension $d$ which is not necessarily a power of two. Quantum information units with $d$ possible states are called qudits. 

The basic quantum operations performed by the quantum gates of the standard toolbox are not always easy to generalize to the qudit setting. In many cases, there is more than one possible generalization, each preserving some of the features of the original qubit gate. 

We present an implementation for a SWAP gate which interchanges two qudits. For two $d$-dimensional states, $\ph$ and $\ps$, the SWAP gate will act as:
\begin{equation}
\mbox{SWAP}\ph\ps=\ps\ph.
\end{equation}
The gate is composed of three equal controlled gates.

\section{Qubit SWAP gates}
\label{qubit}
\begin{figure}[ht!]
\centering
\includegraphics[scale=1]{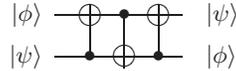}
\caption{CNOT swapping circuit.\label{CNOTSWAP}} 
\end{figure}

An elegant way to implement the SWAP operation for qubits is the CNOT circuit shown in Figure \ref{CNOTSWAP}. The CNOT, or CX, gate is a controlled NOT operation. It has a control qubit, represented as a black dot, and a target qubit, represented with the XOR symbol $\oplus$. If the control qubit is in state $\1$, the target qubit changes its value from $\0$ to $\1$ or from $\1$ to $\0$. The quantum gate operation can be described as 
\begin{equation}
\mbox{CNOT}\ket{x}\ket{y}=\ket{x}\ket{x\oplus y}. 
\end{equation}
The XOR operation is both modulo 2 addition and subtraction. 

This SWAP implementation concatenates three CNOT gates to perform a classical XOR swapping operation in three steps
\begin{equation}
\ket{x}\ket{y}\stackrel{\mbox{\scriptsize{CNOT}}_{2,1}}{\longrightarrow}\ket{x\oplus y}\ket{y},
\end{equation}
\begin{equation}
\ket{x\oplus y}\ket{y}\stackrel{\mbox{\scriptsize{CNOT}}_{1,2}}{\longrightarrow}\ket{x\oplus y}\ket{y\oplus x \oplus y}=\ket{x\oplus y}\ket{x },
\end{equation}
\begin{equation}
\ket{x\oplus y}\ket{x}\stackrel{\mbox{\scriptsize{CNOT}}_{2,1}}{\longrightarrow}\ket{x\oplus y \oplus x}\ket{x}=\ket{y}\ket{x}. 
\end{equation}
Here, $\mbox{CNOT}_{i,j}$ is a CNOT gate controlled by qubit $i$ and with qubit $j$ as target. The evolution we have described is valid for input states $\ket{x}$ and $\ket{y}$ which are $\0$ or $\1$, but general qubit states $\ph$ and $\ps$ can be written as a superposition of these $\0$ and $\1$ states. All the terms in the superposition are swapped and so are the input general states.

\begin{figure}[ht!]
\centering
\includegraphics[scale=1]{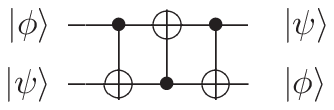}
\caption{Alternative configuration for the CNOT swapping circuit.\label{altCNOTSWAP}} 
\end{figure}

The resulting SWAP circuit works for arbitrary $\ph\ps$ inputs. An input $\ps\ph$ will also be swapped. Due to this symmetry of operation, the alternative ``reflected'' circuit of Figure \ref{altCNOTSWAP} is also a SWAP gate.

\section{The qudit SWAP gate}
\label{qudit}

There are many qudit generalizations of the CNOT operation. Some of them help to build partial swapping circuits \cite{Mer01}, but, in general, they do not offer a complete SWAP gate unless combined with additional gates \cite{Fuj03,PRT09}.  We propose a new qudit SWAP gate with a symmetric configuration that generalizes the circuits of Figures \ref{CNOTSWAP} and \ref{altCNOTSWAP}. The basic building block is what we call the $C\!\widetilde{X}$ gate, an alternative generalization of the CNOT gate which is particularly useful for the SWAP operation. The $C\!\widetilde{X}$ gate and the resulting SWAP circuit will be compared to other generalizations in Section \ref{discussion}.

\subsection{The $C\!\widetilde{X}$ gate}
We define a gate $C\!\widetilde{X}$ acting on qudits $\ket{x}$ and $\ket{y}$ from the basis $\left\{ \ket{0}, \ket{1}, \ldots, \ket{d-1}\right\}$ so that
\begin{equation}
C\!\widetilde{X}\ket{x}\ket{y}=\ket{x}\ket{-x-y}.
\end{equation}
$\ket{-x-y}$ denotes a state $\ket{i}$ in the range $i=0,\ldots,d-1$ with $i=-x-y \mod d$. All the addition and subtraction operations in this paper are considered to be modulo $d$.

When $d=2$, the $C\!\widetilde{X}$ gate performs a CNOT operation. 

\subsection{The qudit SWAP gate}

\begin{figure}[ht!]
\centering
\includegraphics[scale=1]{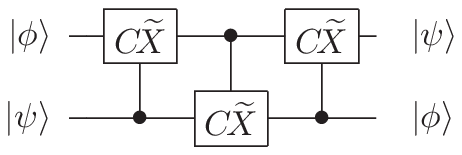}
\caption{Qudit swapping circuit.\label{CXSWAP}} 
\end{figure}

The SWAP gate of Figure \ref{CNOTSWAP} can be extended to qudit inputs using three $C\!\widetilde{X}$ gates (Figure \ref{CXSWAP}). If $C\!\widetilde{X}_{i,j}$ is a $C\!\widetilde{X}$ gate where the control is qudit $i$ and the target qudit $j$, the evolution through the qudit SWAP gate is
\begin{equation}
\ket{x}\ket{y}\stackrel{C\!\widetilde{X}_{2,1}}{\longrightarrow}\ket{-x-y}\ket{y},
\end{equation}
\begin{equation}
\ket{-x-y}\ket{y}\stackrel{C\!\widetilde{X}_{1,2}}{\longrightarrow}\ket{-x-y}\ket{x+y-y}=\ket{-x-y}\ket{x},
\end{equation}
\begin{equation}
\ket{-x-y}\ket{x}\stackrel{C\!\widetilde{X}_{2,1}}{\longrightarrow}\ket{-x+x+y}\ket{x}=\ket{y}\ket{x}.
\end{equation}
The final state is a swapped version of the inputs. Any possible qudit from the input pair is a superposition of states from $\left\{ \ket{0}, \ket{1}, \ldots, \ket{d-1}\right\}$. Therefore, the gate acts as a SWAP for arbitrary input qudits. 

\begin{figure}[ht!]
\centering
\includegraphics[scale=1]{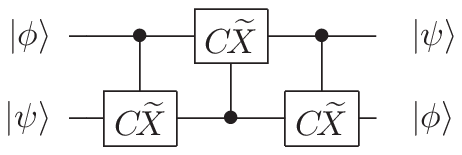}
\caption{Alternative configuration for the qudit swapping circuit.\label{altCXSWAP}} 
\end{figure}

We can also give an alternative circuit inverting the order of the inputs. As the gate is valid for arbitrary inputs, the upside-down circuit of Figure \ref{altCXSWAP} must also perform the SWAP operation.

\subsection{Decomposition of the $C\!\widetilde{X}$ gate in elementary blocks}
The $C\!\widetilde{X}$ gate can be decomposed into three elementary qudit gates, two Quantum Fourier Transform operations, QFT, and the qudit generalization of the $C\!Z$ gate, $C\!Z_d$. The Quantum Fourier Transform is the quantum version of the Discrete Fourier Transform. It takes any qudit state $\ket{x}$ into a uniform superposition 
\begin{equation}
\mbox{QFT}\ket{x}=\frac{1}{\sqrt{d}}\sum_{k=0}^{d-1}e^{i\frac{2\pi x k}{d}}\ket{k}.
\end{equation}

The $C\!Z_d$ gate is a selective phase shift. A control qudit determines the phase shift of the target. Its effect on two input qudits is
\begin{equation}
C\!Z_d\ket{x}\ket{y}=e^{i\frac{2\pi x y}{d}}\ket{x}\ket{y}.
\end{equation}
We also consider its inverse 
\begin{equation}
C\!Z_d^\dag\ket{x}\ket{y}=e^{-i\frac{2\pi x y}{d}}\ket{x}\ket{y}.
\end{equation}

\begin{figure}[ht!]
\centering
\includegraphics[scale=1]{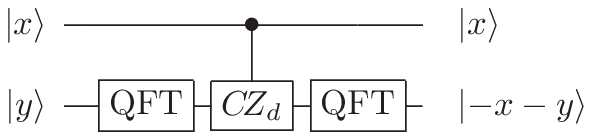}
\caption{Decomposition of the $C\!\widetilde{X}$ gate.\label{CX}} 
\end{figure}

Figure \ref{CX} shows the decomposition of the $C\!\widetilde{X}$ gate in terms of these primitives. We apply a QFT on the target (operation $\mbox{QFT}_2$), then apply a $C\!Z_d$ gate and perform a second $\mbox{QFT}_2$.

The total evolution is the desired $C\!\widetilde{X}$ operation, as
\begin{eqnarray}
\ket{x}\ket{y}&\stackrel{\mbox{\scriptsize{QFT}}_2}{\longrightarrow}&\frac{1}{\sqrt{d}}\sum_{k=0}^{d-1}e^{i\frac{2\pi k y}{d}}\ket{x}\ket{k}\\
&\stackrel{C\!Z_d}{\longrightarrow}&\frac{1}{\sqrt{d}}\sum_{k=0}^{d-1}e^{i\frac{2\pi k y}{d}}e^{i\frac{2\pi x k}{d}}\ket{x}\ket{k}=\frac{1}{\sqrt{d}}\sum_{k=0}^{d-1}e^{i\frac{2\pi k (x+y)}{d}}\ket{x}\ket{k}\\
&\stackrel{\mbox{\scriptsize{QFT}}_2}{\longrightarrow}&\frac{1}{d}\sum_{l=0}^{d-1}\sum_{k=0}^{d-1}e^{i\frac{2\pi k (x+y)}{d}}e^{i\frac{2\pi k l}{d}}\ket{x}\ket{l}=\ket{x}\ket{-x-y}.
\end{eqnarray}
The last step comes from noticing the geometric sum 
\begin{equation}
\sum_{k=0}^{d-1}\left(e^{i\frac{2\pi (x+y+l)}{d}}\right)^k=d\cdot\delta(x+y+l\mod d).
\end{equation}

\begin{figure}[ht!]
\centering
\includegraphics[scale=1]{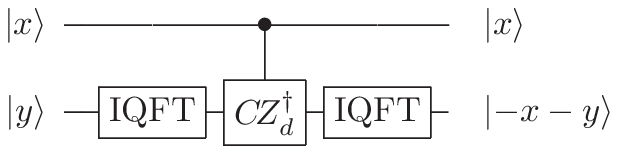}
\caption{Alternative decomposition of the $C\!\widetilde{X}$ gate.\label{CXinv}} 
\end{figure}

We can see the $C\!\widetilde{X}$ gate is its own inverse. If we apply twice the operation $C\!\widetilde{X}\ket{x}\ket{y}=\ket{x}\ket{-x-y}$, we get
\begin{equation}
C\!\widetilde{X}\cdot C\!\widetilde{X}\ket{x}\ket{y}=C\!\widetilde{X}\ket{x}\ket{-x-y}=\ket{x}\ket{-x-(-x-y)}=\ket{x}\ket{y}.
\end{equation}
All quantum gates are unitary operations. If $C\!\widetilde{X}$ is its own inverse, then $C\!\widetilde{X}=C\!\widetilde{X}^\dag$. The decomposition  $C\!\widetilde{X}^\dag=(\mbox{QFT}_2 \cdot C\!Z_d\cdot \mbox{QFT}_2)^\dag=\mbox{IQFT}_2\cdot C\!Z_d^\dag\cdot\mbox{IQFT}_2$ of Figure \ref{CXinv} is also a valid implementation of $C\!\widetilde{X}$.  

Now the elementary gates that give $C\!\widetilde{X}$ are two inverse Quantum Fourier Transforms ($IQFT=QFT^\dag$) and a $C\!Z_d^\dag$ gate in the middle.

\section{Discussion}
\label{discussion}

We have presented an implementation of the SWAP gate valid for states of any arbitrary dimension $d$. The gate preserves some nice properties of the CNOT decomposition of the qubit SWAP. 

Previous qudit SWAP gate proposals use a different generalization of the CNOT gate. The most extended alternative is the $C\!X_d$ gate that transforms states according to
\begin{equation}
C\!X_d\ket{x}\ket{y}=\ket{x}\ket{x+y},
\end{equation}
with a modulo $d$ addition. For qubits, modulo 2 addition and subtraction are the same operation. For qudits, we need a different operation for the inverse. The gate $C\!X_d^\dag$ is given by
\begin{equation}
C\!X_d^\dag\ket{x}\ket{y}=\ket{x}\ket{y-x},
\end{equation}
with a modular subtraction. Combining $C\!X_d$ and $C\!X_d^\dag$ gates, we can give partial SWAP circuits if we know one of the input states. For instance, we can take an input $\ph\0$ into $\0\ph$ \cite{Mer01}. However, these gates are not enough to complete a qudit SWAP gate for any value of $d$ \cite{Wil11,WW12}. The resulting circuits need to be completed with the operation
\begin{equation}
X_d\ket{x}=\ket{d-x}=\ket{-x}
\end{equation}
which offers the modulo $d$ complement of the input. 

\begin{figure}[ht!]
\centering
\includegraphics[scale=1]{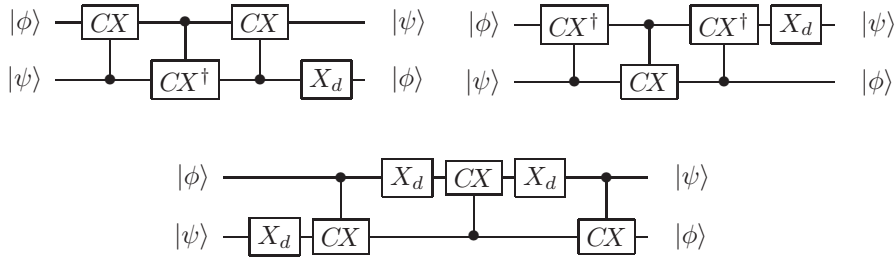}
\caption{Qudit SWAP circuits with the $C\!X_d$, $C\!X_d^\dag$ and $X_d$ gates.\label{asymSWAPs}} 
\end{figure}

The three gates, $C\!X_d$, $C\!X_d^\dag$ and $X_d$, appear in different implementations of the qudit SWAP gate \cite{Fuj03,PRT09}. Figure \ref{asymSWAPs} shows some possible circuits. These gates can be directly adapted to produce a quantum SWAP gate for continuous variables \cite{Wan01,PRT09}. We can similarly give a continuous $C\!\widetilde{X}$ gate.

The qudit SWAP circuit presented in this paper offers a new, simple alternative with only one type of gate. The $C\!\widetilde{X}$ gate we have described is its own inverse and can be decomposed into standard qudit gates.

\begin{acknowledgements}
This work has been funded by MICINN project TEC2010-21303-C04-04.
\end{acknowledgements}
\newcommand{\noopsort}[1]{} \newcommand{\printfirst}[2]{#1}
  \newcommand{\singleletter}[1]{#1} \newcommand{\switchargs}[2]{#2#1}


\begin{thebibliography}{1}
\providecommand{\url}[1]{{#1}}
\providecommand{\urlprefix}{URL }
\expandafter\ifx\csname urlstyle\endcsname\relax
  \providecommand{\doi}[1]{DOI~\discretionary{}{}{}#1}\else
  \providecommand{\doi}{DOI~\discretionary{}{}{}\begingroup
  \urlstyle{rm}\Url}\fi

\bibitem{Mer01}
Mermin, N.D.: From classical state swapping to quantum teleportation.
\newblock Physical Review A \textbf{65}, 012320 (2001).

\bibitem{Fuj03}
Fujii, K.: Exchange gate on the qudit space and Fock space.
\newblock Journal of Optics B: Quantum and Semiclassical Optics \textbf{5}(6),
  S613 (2003).

\bibitem{PRT09}
Paz-Silva, G., Rebi\ifmmode~\acute{c}\else \'{c}\fi{}, S., Twamley, J., Duty,
  T.: Perfect mirror transport protocol with higher dimensional quantum chains.
\newblock Physical Review Letters \textbf{102}, 020503 (2009).

\bibitem{Wil11}
Wilmott, C.: On swapping the states of two qudits.
\newblock International Journal of Quantum Information \textbf{09}(06),
  1511--1517 (2011).

\bibitem{WW12}
Wilmott, C., Wild, P.: On a generalized quantum SWAP gate.
\newblock International Journal of Quantum Information \textbf{10}(03),
  1250034 (2012).

\bibitem{Wan01}
Wang, X.: Continuous-variable and hybrid quantum gates.
\newblock Journal of Physics A: Mathematical and General \textbf{34}, 9577--9584
  (2001).

\end{thebibliography}
\end{document}